%% file: main.tex
\documentclass[a4paper]{article}
\usepackage{ISCSLP2026}
\usepackage{ifthen}

\usepackage[dvipsnames]{xcolor}
\usepackage[pagebackref,breaklinks,colorlinks]{hyperref}
\input{preamble}

\newboolean{blind}
\title{HAMMER: Harmonic-Aware Parallel Context Modeling and Discriminator-Free Perceptual Optimization for Speech Enhancement}

\name{
	\ifthenelse{\boolean{blind}}{Anonymous to ISCSLP}
	{Shang-Fu Chen$^{1,2}$, Szu-Wei Fu$^3$, Sung-Feng Huang$^3$, Rong Chao$^{1,2}$, Wen-Huang Cheng$^{1,4}$, Yu Tsao$^2$ }
}
\address{
  \ifthenelse{\boolean{blind}}{Anonymous to ISCSLP}
  {
  	$^1$National Taiwan University, $^2$Academia Sinica, $^3$Nvidia, $^4$VinUniversity
  }
}

\email{
	\ifthenelse{\boolean{blind}}{Anonymous to ISCSLP}
	{chenshangfu@cmlab.csie.ntu.edu.tw, szuweif@nvidia.com, sungfengh@nvidia.com, roychao19477@gmail.com, wenhuang@csie.ntu.edu.tw, 
    yutsao@as.edu.tw}
}

\begin{document}

\maketitle
\begin{abstract}
Recent speech enhancement systems combine self-attention and Mamba to capture global interactions and long-range dependencies. Yet these hybrids usually operate as sequence mixers and do not explicitly exploit harmonic periodicity, a strong cue for preserving voiced speech under noise. Perceptual optimization poses another challenge. PESQ is non-differentiable, so many methods train auxiliary metric discriminators that increase complexity and introduce adversarial instability. We propose \ours, a harmonic-aware and discriminator-free speech enhancer built around two components. (i) The Time-Frequency Harmonic-aware Attention-Mamba (TF-HAM) block runs self-attention and bidirectional Mamba in parallel along both spectrogram axes, then applies a speech-adapted autocorrelation feed-forward network to encode local periodic structure. (ii) Metric-explicit perceptual refinement (MEPR) combines differentiable PESQ and log-likelihood-ratio losses to expose perceptual metric structure without a learned surrogate. On VoiceBank+DEMAND, \ours achieves 3.69 PESQ and 4.41 COVL with only 2.39\,M parameters, outperforming or matching discriminator-based systems. Inference-time perceptual contrast stretching further raises PESQ to 3.79 without retraining.
The source code will be available at \url{https://github.com/shangfuu/HAMMER.git}.

\end{abstract}

\noindent\textbf{Index Terms}: speech enhancement, speech quality, perceptual loss, state space model, attention mechanism

\input{sections/introduction}

\input{sections/related_work}
\input{sections/method}

\input{sections/experiments}

\input{sections/conclusion}

\bibliographystyle{IEEEtran}
\bibliography{refs}


\end{document}

%% file: preamble.tex
\usepackage{booktabs}
\usepackage{xspace}
\usepackage{comment}
\usepackage{amsmath}
\usepackage{siunitx}    
\usepackage{tabularx}

\newcommand{\ours}{\textbf{HAMMER}\xspace}
\newcommand{\cmark}{$\checkmark$}
\newcommand{\xmark}{$\times$}

\newcommand{\submission}{}

\ifx \submission \undefined
\newcommand{\SC}[1]{\textcolor{Maroon}{#1}}
\else
\newcommand{\SC}[1]{{#1}}
\fi

%% file: sections/introduction.tex
\section{Introduction}

Single-channel speech enhancement (SE) aims to recover clean speech from noisy recordings and supports telephony, hearing aids, and automatic speech recognition. Recent time--frequency systems jointly estimate magnitude and phase in the short-time Fourier transform (STFT) domain. MP-SENet~\cite{lu2023mpsenet} established an effective dual-decoder architecture for this setting, while SEMamba~\cite{chao2024semamba} replaced its Transformer backbone with bidirectional Mamba for compact sequence modeling.
Self-attention and Mamba offer complementary capabilities. Self-attention captures global interactions, whereas Mamba propagates long-range information efficiently through a selective state space. Recent hybrids combine both mechanisms~\cite{zhao2026mambaformer,kim2025mhsenet,kuhne2026mambattention}, but usually stack them sequentially and still act as generic sequence mixers. They model dependencies among time--frequency tokens without explicitly encoding the quasi-periodicity of voiced speech, whose regularly spaced harmonics and repeated temporal patterns remain informative under noise. Representing this structure can help preserve harmonically related speech components while suppressing unstructured interference.


A separate challenge is aligning reconstruction with perceived quality. PESQ is widely used but non-differentiable~\cite{rix2001pesq}, so MetricGAN-style methods~\cite{fu2019metricgan,fu2021metricgan+,cao2022cmgan,chao2024semamba,wang2025zipenhancer,kim2025mhsenet,zhao2026mambaformer} train auxiliary predictors as surrogate objectives, adding complexity and adversarial instability. Differentiable PESQ approximations avoid the discriminator, but replacing hard perceptual operators with smooth alternatives can make their scores deviate from the exact PESQ value, while hard thresholds and piecewise operations can still suppress gradients for near-clean speech~\cite{martin2018pmsqe}. A discriminator-free objective should therefore preserve useful optimization directions without relying on either an exact non-differentiable metric or a learned metric proxy.


Motivated by these observations, we propose \ours (\textbf{H}armonic-aware \textbf{A}ttention-\textbf{M}amba with \textbf{M}etric-\textbf{E}xplicit \textbf{R}efinement), a harmonic-aware and discriminator-free speech enhancer. The proposed \ours have two core components. First, the \emph{Time-Frequency Harmonic-aware Attention-Mamba (TF-HAM)} block couples self-attention and bidirectional Mamba in parallel along both spectrogram axes rather than stacking them sequentially, allowing global token interactions and efficient state-space context to be computed as complementary views. It then applies an autocorrelation feed forward network adapted from Flickerformer~\cite{qu2026flickerformer} to local time--frequency speech patches, bringing periodicity modeling into each block.
Second, we introduce \emph{Metric-Explicit Perceptual Refinement (MEPR)}. MEPR combines a differentiable soft-PESQ loss that turns hard perceptual operators into informative training gradients with a differentiable log-likelihood-ratio (LLR) loss targeting LPC spectral distortion associated with composite speech-quality measures. This objective incorporates perceptual supervision without a metric-prediction discriminator.
On VoiceBank+DEMAND, \ours achieves 3.69 PESQ, 4.83 CSIG, 3.97 CBAK, and 4.41 COVL with only 2.39\,M parameters. It outperforms or matches the discriminator-based systems across all reported metrics. Ablations show complementary gains from parallel hybrid modeling and metric-explicit refinement.

%% file: sections/related_work.tex
\section{Related Work}

\noindent\textbf{Speech Enhancement Models.} MP-SENet~\cite{lu2023mpsenet} established the dual magnitude/phase-decoder codec that most recent systems build on. SEMamba~\cite{chao2024semamba} replaced its Transformer core with bidirectional Mamba, Mamba-SEUNet~\cite{wang2025mambaseunet} scaled the idea into a multi-level U-Net, and hybrids that pair state-space models with attention are emerging in SE~\cite{zhao2026mambaformer}. We adopt the parallel-fusion design of Hymba~\cite{dong2024hymba} from language modelling, merging both mixers inside every time--frequency block. None of these backbones contains a module specialised for harmonicity, which our speech-adapted Autocorrelation
Feed-Forward Network (AFFN) explicitly models.


\input{widget/fig_framework}

\noindent\textbf{Optimizing perceptual metrics.} Because PESQ~\cite{rix2001pesq} is non-differentiable, MetricGAN+~\cite{fu2021metricgan+} and CMGAN~\cite{cao2022cmgan} learn discriminators as metric surrogates, while MetricGAN-OKD~\cite{shin2023metricganokd} extends this idea to multiple metrics. Differentiable surrogates such as PMSQE~\cite{martin2018pmsqe}, PESQ-inspired losses~\cite{kim2019pesqloss}, and torch-pesq~\cite{schmidt2022torchpesq} avoid adversarial training but still require care around hard perceptual operators and metric-specific artifacts~\cite{deoliveira2024pesqetarian}. We instead combine softened PESQ-style supervision with a differentiable LLR term related to composite speech-quality measures~\cite{hu2008evaluation,jin2022composite}.


%% file: widget/fig_framework.tex
\begin{figure*}[ht!]
    \centering
    \includegraphics[width=.98\linewidth]{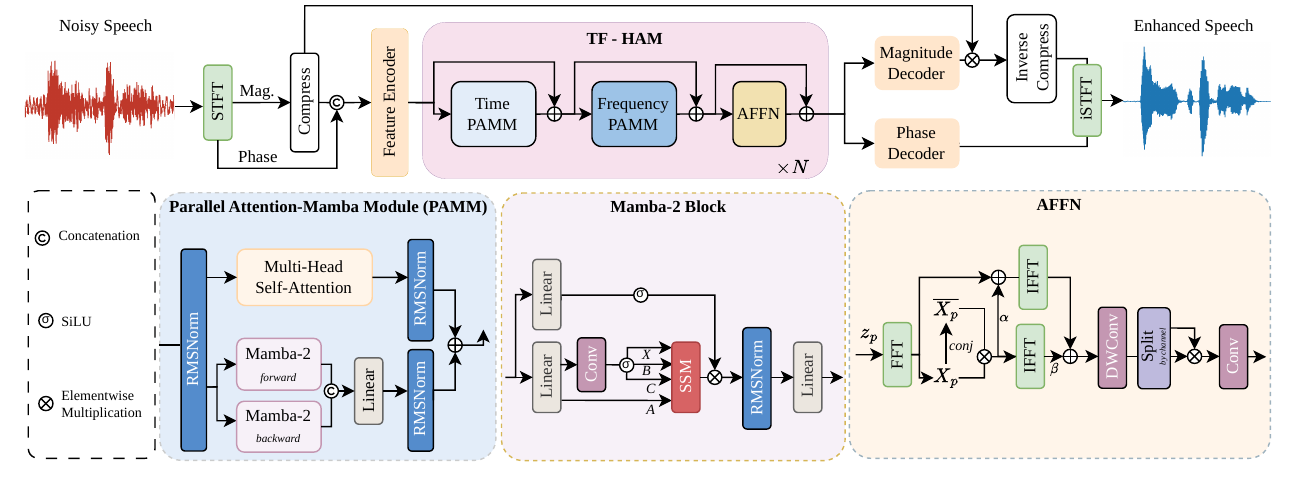}
    \vspace{-1.8em}
    \caption{Overview of \ours. Compressed magnitude and phase spectra are encoded, refined by stacked TF-HAM blocks, and decoded for waveform reconstruction. Each TF-HAM block applies time- and frequency-axis Parallel Attention-Mamba modules followed by a speech-adapted AFFN, combining parallel self-attention, bidirectional Mamba2, and local periodicity modeling.}
    \label{fig:framework}
    \vspace{-1em}
\end{figure*}

%% file: sections/method.tex
\section{Method}

\subsection{Overview}
\SC{As shown in Fig.~\ref{fig:framework}},
\ours follows the complex-spectral, dual-decoder design~\cite{lu2023mpsenet,chao2024semamba,lu2025explicit}. 
Given a noisy waveform, we compute its STFT magnitude and phase, apply power-law magnitude compression, and stack them as a two-channel time--frequency input. A DenseEncoder maps this representation to a compact feature map, which is refined by TF-HAM blocks before separate magnitude and phase decoders reconstruct the enhanced waveform through inverse STFT.
The encoder and decoders are kept unchanged from the SEMamba~\cite{chao2024semamba}, so the proposed design focuses on two parts of the pipeline: (i) the TF-HAM backbone, which combines parallel Attention-Mamba mixing with a speech-adapted autocorrelation feed-forward network (Sec.~\ref{sec:hama}); and (ii) metric-explicit perceptual refinement (MEPR), which combines Soft-PESQ and differentiable LLR without a learned discriminator (Sec.~\ref{sec:losses}).

\subsection{Time-Frequency Harmonic-aware Attention-Mamba}
\label{sec:hama}
Enhancing speech spectrograms requires complementary forms of context. Nonlocal spectral evidence helps recover speech components masked by noise, content-selective long-range propagation supports temporal and spectral continuity, and local periodic structure provides cues for voiced speech. TF-HAM is designed to couple these factors within each block. It first applies a parallel Attention-Mamba mixer along each spectrogram axis, followed by a speech-adapted AFFN. Following the hybrid-head principle of Hymba~\cite{dong2024hymba}, the attention and state-space branches process the same normalized features rather than being stacked sequentially. This avoids imposing a fixed ordering in which one mixer must operate on a representation already filtered by the other. Instead, nonlocal attention and content-selective state-space propagation produce complementary views at the same depth before fusion. The AFFN then provides channel mixing and local periodic-structure modeling.

\noindent\textbf{Parallel Attention-Mamba module.} Given an axis-wise token sequence $x\in\mathbb{R}^{L\times C}$ with $L$ tokens and $C$ channels, TF-HAM applies one pre-normalization and sends the same representation to attention and Mamba2. The attention branch $A$ uses rotary positional embeddings (RoPE)~\cite{su2024roformer}, which rotate queries and keys by their axis positions so that self-attention encodes relative offsets. This branch is computed as:
\begin{equation}
\begin{gathered}
\tilde{x} = \operatorname{RMSNorm}(x), \quad
Q,K,V = \tilde{x}W_{qkv},\\
A = \operatorname{softmax}\!\left(
\frac{\mathcal{R}(Q)\mathcal{R}(K)^{\top}}{\sqrt{d}}\right)V .
\end{gathered}
\end{equation}
where $\mathcal{R}(\cdot)$ denotes the per-head RoPE rotation, $\sqrt{d}$ is the scaling factor, and $W_{qkv}$ denotes linear projection of query, key and value.
In parallel, the Mamba branch $M$ applies bidirectional Mamba2~\cite{dao2024mamba2} to summarize content-selective context:
\begin{equation}
  M =
  [\,\operatorname{SSM}_{\rightarrow}(\tilde{x})\,\|\,\operatorname{SSM}_{\leftarrow}(\tilde{x})\,]W_m,
\end{equation}
where $\|$ denotes channel-wise concatenation, $\operatorname{SSM}_{\rightarrow}$ and $\operatorname{SSM}_{\leftarrow}$ are the forward and backward Mamba2 scans and $W_m$ is a learnable projection.
The outputs of the two branches are then combined to produce:
\begin{equation}
  y = \tfrac{1}{2}\big(\operatorname{RMSNorm}(A)+\operatorname{RMSNorm}(M)\big)W_o.
\end{equation}
Here, $W_o$ is also a learnable linear projection. The mixer output is added to the input by the residual update. This mixer is applied along the temporal and spectral axes in succession, with the residual update performed after each axis-wise pass.

\input{widget/table_experiment}

\noindent\textbf{Speech-adapted AFFN.}
\label{sec:affn}
We adapt the AFFN of Flickerformer~\cite{qu2026flickerformer} to provide channel mixing and a periodic inductive bias for speech spectrograms. While the original module targets image flicker, we apply it to local time--frequency patches, where voiced speech exhibits harmonic repetition. Given features $Z\in\mathbb{R}^{C\times T\times F}$, we first apply a point-wise convolution and partition the normalized features into $P{\times}P$ patches $z_p$ and computes:
\begin{equation}
\begin{gathered}
  \mathcal{X}_p = \mathcal{W}\odot\mathcal{FFT}_{2}(z_p), \quad
  S_p = \mathcal{X}_p\odot\overline{\mathcal{X}_p}, \\
  R_p = \mathcal{FFT}_{2}^{-1}(S_p),
\end{gathered}
\end{equation}
where $p$ indexes a patch, $\mathcal{FFT}_{2}$ is the 2-D FFT, $\mathcal{W}$ is a learnable spectral filter, $\odot$ denotes element-wise multiplication, $\overline{\mathcal{X}_p}$ is the complex conjugate of $\mathcal{X}_p$, $S_p$ is the power spectrum, and $R_p$ is the autocorrelation obtained by the Wiener--Khinchin theorem~\cite{cohen1998generalization}. 
\SC{To make this periodic cue explicit in each local patch, AFFN reinforces periodic structure through:}
\begin{equation}
\begin{gathered}
  \widetilde{z}_p = \mathcal{FFT}_{2}^{-1}(\mathcal{X}_p+\alpha S_p)+\beta R_p,
\end{gathered}
\end{equation}
where $\alpha$ and $\beta$ are learnable scalars.
\SC{The first term can be viewed as a spectrum-enhanced reconstruction. Here, $\mathcal{X}_p$ preserves the learned complex spectrum, while $\alpha S_p$ emphasizes high-power periodic components before mapping back to the patch domain.}
The patches $\{\widetilde{z}_p\}$ are reassembled and passed through gated FFN layers, producing the AFFN output $\widetilde{Z}$. This branch is injected with a zero-initialized residual scale $\gamma$:
\begin{equation}
  Z' = Z+\gamma\,\widetilde{Z}.
\end{equation}

\subsection{Metric-Explicit Perceptual Refinement}
\label{sec:losses}

PESQ-oriented training is often implemented through learned metric discriminators~\cite{fu2021metricgan+,cao2022cmgan,chao2024semamba,wang2025zipenhancer,kim2025mhsenet,zhao2026mambaformer}, because the reference PESQ implementation is non-differentiable.
This introduces an extra network, additional computation, and adversarial training instability. MEPR instead keeps the standard losses as the training anchor and adds two direct differentiable metric losses: Soft-PESQ for perceptual disturbance and log-likelihood-ratio (LLR) loss for linear predictive coding (LPC) spectral distortion. Both are computed from clean and enhanced waveforms, without a learned discriminator.

\noindent\textbf{Soft-PESQ.} Although the official PESQ implementation is not differentiable, most of its perceptual pipeline consists of deterministic signal-processing operations, such as filtering, time--frequency transforms, Bark-scale mapping, loudness conversion, and disturbance aggregation, that can be written as tensor operations with autograd.
Such a graph is a training surrogate rather than the exact evaluation metric: its absolute score may differ from the reference PESQ implementation, but its perceptual disturbance signal is still useful for guiding enhancement.
\SC{However, directly differentiating this surrogate still inherits hard dead-zones, thresholds, and saturation operations from the original PESQ pipeline, which can yield zero or abrupt gradients and cause learning to stall or become unstable.}
We therefore start from the differentiable P.862-style~\cite{schmidt2022torchpesq} and soften the hard operators that create gradient dead zones or discontinuities.

Inside the differentiable PESQ pipeline, a non-smooth step is the dead-zone applied to each signed loudness disturbance $d$ with masking threshold $z$. Instead of setting all sub-threshold disturbances to zero, we use a leaky dead-zone:
\begin{equation}
  \widetilde{d}=\operatorname{sgn}(d)
  \left[\max(|d|-z,0)+\lambda\min(|d|,z)\right],
\end{equation}
where $\operatorname{sgn}(\cdot)$ returns the sign of its argument, and $\lambda{=}0.05$ keeps a small gradient inside the dead-zone.
We apply the same principle to the other hard decisions: the asymmetry threshold on the disturbance ratio $r$ is replaced by a sigmoid gate $\sigma((r-3)/\tau_g)$ with $\tau_g{=}0.5$, and each hard upper cap $\min(u,c)$ is replaced by $u-\tau_s\log(1+\exp((u-c)/\tau_s))$, where $u$ is the input, $c$ is the saturation limit, and $\tau_s{=}2$ controls the transition smoothness.
The softened PESQ pipeline directly computes the Soft-PESQ loss $\mathcal{L}_{\text{pesq}}$ from clean waveform $x$ and enhanced waveform $\hat{x}$, using a loss scaling factor of $0.5$.
Other numerical stabilizers remain unchanged.

\noindent\textbf{Differentiable LLR.} Soft-PESQ measures perceptual disturbance after auditory-domain processing, but it does not explicitly constrain the speech spectral envelope. We therefore add a log-likelihood ratio (LLR) term~\cite{hu2008evaluation} on linear predictive coding (LPC) spectra, which penalizes mismatches in the all-pole envelope that captures formant structure and vocal-tract coloration.
We implement the LLR computation with autograd-compatible tensor operations. Clean and enhanced waveforms are framed at 16\,kHz using 480-sample windows and 120-sample hops, multiplied by a Hann window, and converted to order-$P{=}16$ autocorrelations. A batched Levinson--Durbin recursion, following the same reference implementation, produces LPC coefficient vectors $a_x, a_{\hat{x}} \in \mathbb{R}^{P+1}$ (including the leading unity tap) for the clean and enhanced frames, respectively, and the frame-level distortion is:
\begin{equation}
  d_\text{LLR}
  = \log\frac{a_{\hat{x}}^{\top} R_x\, a_{\hat{x}} + \epsilon}
             {a_x^{\top} R_x\, a_x + \epsilon},
\end{equation}
where $R_x \in \mathbb{R}^{(P+1)\times(P+1)}$ is the Toeplitz autocorrelation matrix of the clean frame and $\epsilon = 10^{-10}$ stabilizes the logarithm. As in the composite objective measure~\cite{hu2008evaluation}, we sort frame-level distortions per utterance, keep the lowest $95\%$, and average them to obtain $\mathcal{L}_\text{llr}$. Sorting acts as an index permutation with a well-defined subgradient almost everywhere, so the truncated average remains differentiable with respect to the enhanced waveform.

\noindent\textbf{Total objective.} We follow the SEMamba reconstruction objective~\cite{chao2024semamba} and add the Soft-PESQ and differentiable LLR losses:
\begin{align}
  \mathcal{L} =\;& 0.9\,\mathcal{L}_\text{mag} + 0.3\,\mathcal{L}_\text{pha}
  + 0.1\,\mathcal{L}_\text{com} + 0.2\,\mathcal{L}_\text{time} \nonumber\\
  &+ 0.1\,\mathcal{L}_\text{con} + 0.2\,\mathcal{L}_\text{pesq} + 0.1\,\mathcal{L}_\text{llr},
\end{align}
where $\mathcal{L}_\text{mag}$, $\mathcal{L}_\text{pha}$, $\mathcal{L}_\text{com}$, $\mathcal{L}_\text{time}$, and $\mathcal{L}_\text{con}$ are the SEMamba reconstruction losses, and $\mathcal{L}_\text{pesq}$ and $\mathcal{L}_\text{llr}$ are the metric-explicit terms defined above. No adversarial term is used.


%% file: widget/table_experiment.tex
\begin{table*}[ht!]
\centering
\caption{Comparison of speech enhancement models on VoiceBank+DEMAND. Best value per column in \textbf{bold}; \underline{underline} marks our best result where a baseline holds the column best.}
\vspace{-1em}
\label{tab:comparison}
\begin{tabularx}{0.95\textwidth}{@{\extracolsep{\fill}}llcccccc}
\toprule
Model & Venue & PESQ & CSIG & CBAK & COVL & STOI & Params \\
\midrule
Noisy & -- & 1.97 & 3.35 & 2.44 & 2.63 & 0.92 & -- \\
\midrule
SEGAN~\cite{pascual2017segan} & Interspeech 2017 & 2.16 & 3.48 & 2.94 & 2.80 & -- & 43.18M \\
Demucs~\cite{defossez2020demucs} & Interspeech 2020 & 3.07 & 4.31 & 3.40 & 3.63 & 0.95 & 33.53M \\
MetricGAN+~\cite{fu2021metricgan+} & Interspeech 2021 & 3.15 & 4.14 & 3.16 & 3.64 & 0.93 & -- \\
SE-Conformer~\cite{kim2021seconformer} & Interspeech 2021 & 3.13 & 4.45 & 3.55 & 3.82 & 0.95 & -- \\
TSTNN~\cite{wang2021tstnn} & ICASSP 2021 & 2.96 & 4.33 & 3.53 & 3.67 & 0.95 & 0.92M \\
DPT-FSNet~\cite{kim2022dpt} & ICASSP 2022 & 3.33 & 4.58 & 3.72 & 4.00 & 0.96 & -- \\
DPCFCS-Net~\cite{wang2023dpcfcs} & Interspeech 2023 & 3.42 & 4.71 & 3.88 & 4.15 & 0.96 & 2.86M \\
S4ND-UNet~\cite{saric2024s4ndunet} & Interspeech 2023 & 3.15 & 4.52 & 3.62 & 3.85 & -- & 0.75M \\
MP-SENet~\cite{lu2023mpsenet} & Interspeech 2023 & 3.50 & 4.73 & 3.95 & 4.22 & 0.96 & 2.05M \\
MUSE~\cite{lin2024muse} & Interspeech 2024 & 3.37 & 4.63 & 3.80 & 4.10 & 0.95 & 0.51M \\
CMGAN~\cite{cao2022cmgan} & T-ASLP 2024 & 3.41 & 4.63 & 3.94 & 4.12 & 0.96 & 1.83M \\
SEMamba~\cite{chao2024semamba} & SLT 2024 & 3.55 & 4.77 & 3.95 & 4.29 & 0.96 & 2.25M \\
ZipEnhancer (S, $\lambda_6=0.2$)~\cite{wang2025zipenhancer} & ICASSP 2025 & 3.61 & 4.81 & 3.97 & 4.35 & 0.96 & 2.04M \\
Mamba-SEUNet (M)~\cite{wang2025mambaseunet} & ICASSP 2025 & 3.57 & 4.79 & 4.00 & 4.30 & 0.96 & 3.78M \\
Mamba-SEUNet (S)~\cite{wang2025mambaseunet} & ICASSP 2025 & 3.54 & 4.77 & 3.98 & 4.28 & 0.96 & 1.88M \\
MH-SENet~\cite{kim2025mhsenet} & Interspeech 2025 & 3.62 & \underline{4.79} & \underline{4.01} & 4.34 & 0.96 & 0.99M \\
Mamba-Former (S, $\lambda_6=0.2$)~\cite{zhao2026mambaformer} & ICASSP 2026 & \underline{3.64} & \textbf{4.83} & \textbf{4.02} & \underline{4.39} & 0.96 & 2.14M \\
\midrule
\textbf{HAMMER} & -- & \textbf{3.69} & \textbf{4.83} & 3.97 & \textbf{4.41} & 0.96 & 2.39M \\
\bottomrule
\end{tabularx}
\vspace{-1em}
\end{table*}

%% file: sections/experiments.tex
\section{Experiments}

\input{widget/table_ablation}

\subsection{Experimental Detail}
\noindent \textbf{Dataset.} We evaluate on VoiceBank+DEMAND~\cite{valentini2016vctk}, the standard single-channel benchmark. The training set pairs $11{,}572$ utterances from $28$ speakers with DEMAND noise at signal-to-noise ratios of $\{0,5,10,15\}$\,dB, and the test set contains $824$ utterances from $2$ unseen speakers mixed at $\{2.5,7.5,12.5,17.5\}$\,dB. All audio is resampled to $16$\,kHz.

\noindent \textbf{Configuration.} STFT features use a 400-sample Hann window, 100-sample hop, $n_\text{fft}{=}400$, and power-law magnitude compression with $c{=}0.3$. We use $C{=}64$ channels, $N{=}4$ TF-HAM blocks, $H{=}4$ attention heads, and Mamba2 settings $d_\text{state}{=}16$, $d_\text{conv}{=}4$, $\text{expand}{=}4$, totaling $2.39$\,M parameters. We train 2-second segments with AdamW ($\beta_1{=}0.8$, $\beta_2{=}0.99$, learning rate $5\!\times\!10^{-4}$ decayed by $0.99$ per epoch), bf16 mixed precision, and batch size $4$ per GPU on two GPUs. Perceptual-loss models use gradient clipping at norm $1.0$ and an EMA generator (decay $0.999$), converging within $70$ epochs. Following prior work~\cite{chao2024semamba}, we also report perceptual contrast stretching (PCS)~\cite{chao2022pcs}, applied at inference rather than retraining.


\noindent \textbf{Evaluation Metrics.} We report wide-band PESQ, the composite predictors CSIG, CBAK, and COVL~\cite{hu2008evaluation} and STOI. Higher is better for all. Enhanced utterances are decompressed and inverse-transformed before scoring with the reference implementations, so the numbers are directly comparable to published results.

\subsection{Comparison with State of the Art}
Table~\ref{tab:comparison} compares \ours with representative time-domain, GAN-based, Transformer, state-space, and sequential hybrid attention-Mamba systems. \ours attains $3.69$ PESQ and $4.41$ COVL with only $2.39$\,M parameters, giving the best overall quality while remaining close to the strongest background-quality scores. Notably, this performance is achieved without a metric discriminator, indicating that direct differentiable perceptual optimization can be competitive with adversarial metric prediction. Together, the results show that \ours offers a favorable quality--complexity trade-off among compact speech-enhancement models.

\input{widget/table_pcs}

\subsection{Ablation Study}
Table~\ref{tab:ablation} isolates the contribution of TF-HAMA and MEPR using SEMamba as the closest baseline. Soft-PESQ alone provides a modest but consistent gain, indicating that direct perceptual gradients are useful even without changing the backbone. Replacing the backbone with TF-HAMA yields the largest improvement, suggesting that harmonic-aware parallel Mamba-attention contributes more than simply adding a perceptual loss. Adding LLR further improves CSIG and COVL, consistent with its role in constraining LPC spectral distortion rather than directly maximizing PESQ. 
A sequential Mamba-to-attention variant with the same losses obtains 3.67 PESQ and 4.40 COVL, slightly below TF-HAMA, supporting parallel fusion over fixed-order stacking.
Overall, the ablation shows that architecture and optimization are complementary: TF-HAMA improves the representation, while MEPR refines it toward perceptual quality.


\subsection{Inference-Time Perceptual Contrast Stretching}
\label{sec:pcs}
Prior systems obtain PCS gains by \emph{retraining} on PCS-processed targets, fixing one PESQ--CBAK trade-off in the weights. We instead apply PCS directly at inference, using $\alpha$ to control the operating point. Table~\ref{tab:pcs} shows that increasing $\alpha$ improves PESQ/COVL up to moderate strengths but monotonically reduces CBAK. Thus, one \ours checkpoint traces multiple quality--background trade-offs, reaching $3.79$ PESQ at $\alpha{=}0.65$ or the strongest CBAK ($3.87$) at $\alpha{=}0.3$, without retraining.

%% file: widget/table_ablation.tex
\begin{table}[t]
\centering
\caption{Ablation study on VoiceBank+DEMAND. MEPR is decomposed into soft-PESQ and LLR.}
\vspace{-1em}
\label{tab:ablation}
\resizebox{\linewidth}{!}{%
\begin{tabular}{ccccccc}
\toprule
 TF-HAMA & Soft-PESQ & LLR & PESQ & CSIG & CBAK & COVL \\ 
\midrule
\xmark & \xmark & \xmark & 3.55 & 4.77 & 3.95 & 4.29 \\
\xmark & \cmark & \xmark & 3.58 & 4.77 & 3.98 & 4.31 \\
\cmark & \cmark & \xmark & \textbf{3.69} & 4.81 & \textbf{4.00} & 4.39 \\
\cmark & \cmark & \cmark & \textbf{3.69} & \textbf{4.83} & 3.97 & \textbf{4.41} \\
\bottomrule
\end{tabular}
}
\vspace{-1.5em}
\end{table}

%% file: widget/table_pcs.tex
\begin{table}[t]
\centering
\caption{Perceptual contrast stretching (PCS) on VoiceBank+DEMAND. Baselines use PCS retraining, while \ours applies PCS at inference. Best values are in \textbf{bold}.}
\vspace{-1em}
\label{tab:pcs}
\setlength{\tabcolsep}{4pt}
\resizebox{\linewidth}{!}{
\begin{tabular}{lccccc}
\toprule
Model & PESQ & CSIG & CBAK & COVL & STOI \\
\midrule
SEMamba +PCS~\cite{chao2024semamba} & 3.69 & 4.79 & 3.63 & 4.37 & \textbf{0.96} \\
Mamba-SEUNet (S) +PCS~\cite{wang2025mambaseunet} & 3.70 & 4.79 & 3.64 & 4.37 & \textbf{0.96} \\
Mamba-SEUNet (L) +PCS~\cite{wang2025mambaseunet} & 3.73 & 4.82 & 3.67 & 4.40 & \textbf{0.96} \\
\midrule
\ours + PCS($\alpha{=}0.3$)  & 3.75 & 4.84 & \textbf{3.87} & 4.46 & 0.95 \\
\ours + PCS($\alpha{=}0.5$)  & 3.78 & \textbf{4.85} & 3.80 & \textbf{4.47} & 0.95 \\
\ours + PCS($\alpha{=}0.65$) & \textbf{3.79} & \textbf{4.85} & 3.76 & \textbf{4.47} & 0.95 \\
\ours + PCS($\alpha{=}1.0$)  & 3.77 & 4.83 & 3.67 & 4.45 & 0.95 \\
\bottomrule
\end{tabular}
}
\vspace{-1.5em}
\end{table}

%% file: sections/conclusion.tex
\section{Conclusion}
We presented \ours, a 2.39\,M-parameter speech enhancer that addresses two common forms of indirection in modern SE, namely generic sequence mixing and learned metric prediction. TF-HAM brings harmonic-aware parallel Attention-Mamba and local periodicity modeling into each time--frequency block, while MEPR turns perceptual metric structure into direct differentiable supervision without an auxiliary discriminator. On VoiceBank+DEMAND, \ours reaches $3.69$ PESQ and $4.41$ COVL ($3.79$ PESQ with inference-time PCS), comparing favorably with larger systems and adversarial metric optimizers. Ablations confirm complementary gains from the proposed components, and a single checkpoint supports a controllable PESQ--CBAK trade-off through the PCS strength $\alpha$. These results suggest that speech enhancement can benefit from making both speech structure and perceptual objectives explicit, rather than asking generic backbones and learned proxies to discover them implicitly.
